\documentstyle[12pt,epsf,amssymb]{article}

\begin{document}

\baselineskip 6mm
\renewcommand{\thefootnote}{\fnsymbol{footnote}}


\newcommand{\nc}{\newcommand}
\newcommand{\rnc}{\renewcommand}


\rnc{\baselinestretch}{1.24}    
\setlength{\jot}{6pt}       
\rnc{\arraystretch}{1.24}   

\makeatletter
\rnc{\theequation}{\thesection.\arabic{equation}}
\@addtoreset{equation}{section}
\makeatother



\nc     {\be}                   {\begin{equation}}
\nc     {\ee}                   {\end{equation}}
\nc     {\bea}                  {\begin{eqnarray}}
\nc     {\eea}                  {\end{eqnarray}}
\nc     {\ba}                   {\begin{array}}
\nc     {\ea}                   {\end{array}}

\nc     {\nn}                   {\nonumber\\}
\nc     {\ct}                   {\cite}
\nc     {\la}                   {\label}
\nc     {\eq}[1]                {Eq.(\ref{#1})}


\rnc    {\c}[1]     {{\cal #1}}


\nc     {\id}       {{\mathbb I}}


\def    \IR         {{\hbox{{\rm I}\kern-.2em\hbox{\rm R}}}}
\def    \IB         {{\hbox{{\rm I}\kern-.2em\hbox{\rm B}}}}
\def    \IN         {{\hbox{{\rm I}\kern-.2em\hbox{\rm N}}}}
\def    \IC         {\,\,{\hbox{{\rm I}\kern-.59em\hbox{\bf C}}}}
\def    \IZ         {{\hbox{{\rm Z}\kern-.4em\hbox{\rm Z}}}}
\def    \IP         {{\hbox{{\rm I}\kern-.2em\hbox{\rm P}}}}
\def    \IH         {{\hbox{{\rm I}\kern-.4em\hbox{\rm H}}}}
\def    \ID         {{\hbox{{\rm I}\kern-.2em\hbox{\rm D}}}}


\def    \a          {\alpha}
\def    \b          {\beta}
\def    \g          {\gamma}
\def    \d          {\delta}
\def    \ep         {\epsilon}
\def    \ph         {\phi}
\def    \k          {\kappa}
\def    \l          {\lambda}
\def    \m          {\mu}
\def    \n          {\nu}
\def    \th         {\theta}
\def    \rh         {\rho}
\def    \s          {\sigma}
\def    \S          {\Sigma}
\def    \t          {\tau}
\def    \om         {\omega}
\def    \G          {\Gamma}
\def    \L          {\Lambda}
\def    \Om         {\Omega}


\def    \pr         {\prime}
\def    \p          {\partial}
\def\half{\frac{1}{2}}
\def\lint#1#2{\int\limits_{#1}^{#2}}
\def\lsum#1#2{\sum\limits_{#1}^{#2}}
\def\goto{\rightarrow}
\def\para{\parallel}

\def\bk#1{\langle #1 \rangle}        
\def\bkb#1{\left[  #1 \right]}       
\def\bkm#1{\left\{ #1 \right\}}      
\def\bks#1{\left(  #1 \right)}       
\nc     \ltb     {\left[}       
\nc     \ltm     {\left\{}       
\nc     \lts     {\left(}       
\nc     \rte     {\right.}
\nc     \rtb     {\right]}
\nc     \rtm     {\right\}}
\nc     \rts     {\right)}
\nc     \lte     {\left.}


\def\Tr{{\rm Tr}\,}
\def\det{{\rm det}}



\def\so{SO(4)}
\def\sop{SO(4)^\prime}
\def\bc{{\bf C}}
\def\bfz{{\bf Z}}
\def\bz{\bar{z}}

\begin{titlepage}

\hfill\parbox{5cm} { }

\vspace{25mm}

\begin{center}
{\Large \bf  Open String Spectrum in pp-wave background}

\vspace{15mm}
Chan Yong Park$^{\, a \,}$\footnote{cyong21@hepth.sogang.ac.kr}
\\[10mm]

${}^a$ {\sl Department of Physics, Sogang University,
Seoul 121-742, Korea} \\
\end{center}

\thispagestyle{empty}

\vskip2cm


\centerline{\bf ABSTRACT}
\vskip 4mm

\noindent
In this paper, we consider the D-brane, especially ${\rm D}_-$-brane,
in the pp-wave background, which has
the eight dynamical and the eight kinematical supercharges. Since
the pp-wave background has not a SO(8) but a ${\rm SO}(4) \times {\rm SO}(4)$
symmetric group, the D-brane world-volume theory has a non-trivial symmetric
group which depends on the configuration of D-brane. Here, we will analyze the
open string spectrum consistent with the non-trivial symmetric group and, at the
low energy limit, classify the field contents of the D-brane world-volume theory which
come from the fermionic zero modes of the open string.


\vspace{2cm}

\today

\end{titlepage}

\renewcommand{\thefootnote}{\arabic{footnote}}
\setcounter{footnote}{0}

\section{Introduction}

Recently, in Ref. \ct{blau} it was shown that the pp-wave
background is a maximal supersymmetric space like Minkowski,
anti-de Sitter(AdS) and de Sitter(dS) space and that the closed
string theory on this pp-wave background with the RR five-form
flux is exactly solvable \ct{met1,met2} in the light-cone gauge.
Moreover, this pp-wave background can be obtained from $AdS_5
\times S^5$ by taking the Penrose limit. In according to the
AdS/CFT correspondence, it was suggested that the closed string
theory in this background has a similar correspondence with the
gauge theory with the large R-charge \ct{bmn}. By many authors it
was shown that this conjecture including an interaction terms in
the gauge theory \ct{kris}-\ct{gursoy} side and the superstring
side \ct{kiem1}-\ct{kiem2} is still valid. In ref. \ct{spradlin},
the closed superstring field theory is constructed with the
similar method used in flat space \ct{gs1,gs2,gs3} and the
interaction hamiltonians for the closed \ct{spradlin} and open \ct{ck} string theories
are determined by requiring the closure of
the pp-wave background superalgebra .

Furthermore, D-branes in the pp-wave background described by an open string, which
is obtained from the type IIB closed string using the boundary conditions \ct{lam,dab},
was identified and their supersymmetries was classified \ct{billo}-\ct{hsy1}.
In addition, supersymmetries
of the intersection branes is also investigated \ct{hsy2,ohta}.
In Ref. \ct{sken1,sken2}, it was shown that there can be
two kinds of D-branes, for example ${\rm D}_-$-branes, ${\rm D}_+$-branes,
in pp-wave background. These two kinds of branes appear due
to the non-trivial open string boundary condition of the mass term caused by the
pp-wave metric. Moreover, since the symmetric group of this pp-wave background is
given by ${\rm SO(4)} \times {\rm SO(4)}$ due to the existence of the five-form flux,
each brane having a special configuration exists.
This kind of a non-trivial configuration of D-brane induces the non-trivial symmetric
group structure of the world-volume theory on the D-brane in the pp-wave background
\ct{park1}.
Usually, in the flat space background the world-volume theory on D-brane becomes SYM (super
Yang-Mills) theory which corresponds to the open string theory, especially the fermionic
zero modes, at the low energy limit.
To describe the world-volume theory having a non-trivial symmetric group,
in the first place we consider the open string theory in the pp-wave background and then
find the open string spectrum at the low energy limit which becomes field contents of
the SYM theory with a non-trivial symmetric group.

In this paper, we will explicitly show that, especially in the ${\rm D}_-$-brane cases,
how we can obtain the field contents of world-volume theory on D-brane from
the open string theory in the pp-wave background.
The paper is organized as follows. In Sec. 2, we will give a brief review for the open string
theory in the pp-wave. In Sec. 3, using the open string spectrum we will show that the
field contents of the world-volume theory are obtained from the open string theory at the
low energy limit. In Sec. 4, we will conclude this paper with some comments. In the
appendix, we will give a shot explanation for the spinor representation.

\section{Review of open string in pp-wave background}

In the pp-wave background including a constant R-R five-form
flux, the metric is given by \ct{hsy2}
\be
    ds^2 = 2 dx^+ dx^- - \mu^2 (x^I)^2 (dx^+)^2
+ \delta_{IJ} dx^I dx^J ,
\ee
and
\be
F_{+1234} = F_{+5678} = 2 \mu ,
\ee
where $\mu$ is an arbitrary constant and index $I$ runs
from $1$ to $8$. These quantities can be obtained by maximally
supersymmetric solutions of the type IIB string theory.

The Green-Schwarz light-cone action in the plane wave background
describe massive bosons and fermions. In the light-cone gauge,
$X^+ = \tau$, the action is given by
\be   \la{act1}
    S = \frac{1}{2\pi \alpha^{\prime} p^+} \int d\tau {\int_0}^{2\pi
        \alpha^{\prime} |p^+ |}
        d\sigma \left[ \frac{1}{2} \partial_+ X_I \partial_- X_I
        -\frac{1}{2} \mu^2 {X_I}^2
        - i \bar{S} (\rho^A \partial_A - \mu \Pi) S \right]
\ee
where $\partial_\pm = \partial_{\tau} \pm \partial_{\sigma}$.
In this paper, we will set $\alpha = 2 \alpha^{\prime} p^+$ for
open strings and take the spinors $S$ as eight two-components
Majorana spinors on the world-sheet $\Sigma$ that transform as
positive chirality spinors ${\bf 8}_s$ under $SO(8)$:
\be
S^a= \left(%
\begin{array}{c}
  S^{1a} \\
  S^{2a} \\
\end{array}%
\right), \qquad
\bar{S}^a= {S^a}^T \rho^\tau,
\ee
where
\be \la{2d-dirac}
\rho^\tau=\left(%
\begin{array}{cc}
  0 & -1 \\
  1 & 0 \\
\end{array}%
\right), \qquad
\rho^\sigma =\left(%
\begin{array}{cc}
  0 & 1 \\
  1 & 0 \\
\end{array}%
\right).
\ee
The presence of $\Pi$ in the fermionic action breaks the SO(8) symmetry into
${\rm SO(4)} \times {\rm SO(4)}^{\pr}$. From this action \eq{act1} we can
obtain the equations of motion of the bosonic field $X^I$ and the fermionic field
$S^a$:
\bea \la{eom-boson}
&& \p_+\p_-X^I + \mu^2 X^I =0, \\
\la{eom-fermion}
&& \p_+ S^1 - \mu \Pi S^2 =0, \qquad \p_-S^2 + \mu \Pi S^1 =0.
\eea

\subsection{The bosonic sectors}

The Green-Schwarz action for bosonic modes is given by
\be
S_B = \frac{1}{\pi \alpha} \int d\tau {\int_0}^{\pi |\alpha|} d \sigma
\left[ \frac{1}{2} \partial_+ X_I \partial_- X_I -\frac{1}{2} \mu^2 {X_I}^2
\right] .
\ee
Now we pay attention to open strings ending on Dp-branes, especially static Dp-branes
for simplest, in plane wave background. For these open strings, Neumann
boundary conditions,
\be
\partial_{\sigma} X^r |_{\partial \Sigma} = 0 ,
\ee
are imposed along directions of the Dp-branes world-volume and Dirichlet
boundary conditions,
\be
\partial_{\tau} X^{r^{\prime}} |_{\partial \Sigma} = 0,
\ee
for remaining transverse coordinates where $X^r$ and $X^{r^{\prime}}$
imply the longitudinal and
transverse coordinates on Dp-branes respectively. The solutions satisfying
equations on motion and
boundary conditions are given by
\bea
X^r (\sigma, \tau) &=& x_0 ^r \cos \mu \tau + \frac{p_0 ^r}{\mu}
                     \sin \mu \tau
                     + i \sum_{n \ne 0} \frac{1}{\omega_n}
                     \alpha_n ^r e^{-i\omega_n \tau}
                     \cos \frac{n \sigma}{|\alpha|} , \nn
X^{r^{\prime}} (\sigma, \tau) &=& x_0 ^{r^{\prime}} (\sigma)
                     + \sum_{n \ne 0} \frac{1}{\omega_n}
                     \alpha_n ^{r^{\prime}}
                     e^{-i\omega_n \tau} \sin \frac{n\sigma}{|\alpha|} ,
\eea
where the zero mode parts, $x_0 ^{r^{\prime}} (\sigma)$, represent the
positions of D-brane in
transverse directions and are given by
\be
x_0 ^{r^{\prime}} (\sigma) = \frac{x_0 ^{r^{\prime}}}{1+e^{\mu |\alpha|
                                \pi}}
                              \left( e^{\mu \sigma} + e^{\mu (|\alpha| \pi - \sigma} \right) .
\ee
Using the canonical conjugate momentum defined by
\be
P_I \equiv \frac{\partial {\cal L}}{ \partial \partial_{\tau} X^I}  \; ,
\ee
the Hamiltonian is given by
\be
    \c{H}_B = \frac{1}{ 2 \pi \alpha} \left[ (\pi \alpha)^2 {P_I}^2
           + (\partial_{\sigma} )^2 + \mu^2 {X_I}^2 \right] ,
\ee
where two phase space coordinates, $X^I$ and $P_I$, satisfy the following
commutation relation
\be
\left[ X^I (\sigma) , P^J (\sigma^{\prime}) \right]  = i \delta^{IJ}
                   \delta(\sigma-\sigma^{\prime}) .
\ee
From the above commutation relation, we can get commutation relations of
expanding modes
\be
\left[ x_0 ^r , p_0 ^s \right] = i \delta^{rs} \;\; {\rm and} \;\;
     \left[ \alpha_n ^I , \alpha_m ^J \right] = \omega_n \delta_{n+m, 0}
              \delta^{IJ}  ,
\ee
where $x_0 ^r$ and $p_0 ^s$ are zero modes of $X^r$ and $P^s$
respectively and
\be
\omega_n = {\rm sign}(n) \sqrt{\mu^2 + n^2 / \alpha^2} .
\ee
In terms of modes $n$, the Hamiltonian is given by
\be     \label{bosham}
    \c{H}_B = {\rm sign}(\alpha) \left[ \frac{1}{2} {p_0 ^r}^2
            + \frac{1}{2} {x_0 ^r}^2
            + \frac{1}{2} \sum_{n \ne 0} {\alpha_n}^I {\alpha_{-n}}^I \right] .
\ee
By introducing new variables for $n=0$
\bea
a_0 ^{r} &=& \frac{1}{\sqrt{2}} \left( \frac{p_0 ^r}{\sqrt{\omega_0}}
             - i \sqrt{\omega_0} {x_0 ^r} \right) , \nn
{a_0 ^r}^+ &=& \frac{1}{\sqrt{2}} \left( \frac{p_0 ^r}{\sqrt{\omega_0}}
             + i \sqrt{\omega_0} {x_0 ^r} \right) ,
\eea
and for $n > 0$
\bea
a_n ^I &=& \frac{\alpha_n ^I}{\sqrt{\omega_n }} , \nn
{a_n ^I}^+ &=& \frac{\alpha_{-n} ^I}{\sqrt{\omega_n }} ,
\eea
this Hamiltonian can be rewritten as
a harmonic oscillator form
\be
    \c{H}_B = {\rm sign} (\alpha)
                \frac{1}{2} \sum_{n \ge 0} \omega_n
           \left( {a_n ^I}^+ a_n ^I + a_n ^I {a_n ^I}^+ \right) ,
\ee
satisfying the commutation relations of simple canonical form
\be
\left[ a_0 ^r , {a_0 ^s}^+ \right] = \delta^{rs} \;\; {\rm and} \;\;
     \left[ a_n ^I , {a_m ^J}^+ \right] = \delta^{IJ} \delta_{n,m} .
\ee
Notice that $a_0 ^{r^{\pr}}$ and $ {a_0 ^{s^{\pr}}}^+$, which are
zero modes
of the transverse coordinates, are set to zero. Since the energy
contributions of zero modes of
the transverse coordinates are absent, the zero point energy of the above
hamiltoian is given by
\be     \label{boszero}
E_{B,0} = {\rm sign} (\alpha) \cdot \frac{(p-1)\omega_0}{2},
\ee
where $(p-1)$ is the number of coordinates satisfying the Neunmann boundary
condition except light-cone coordinates. This
\eq{boszero} implies that the bosonic zero point energy depends on a world
volume dimension of
Dp-brane.

\subsection{The fermionic sectors}

The action for fermionic field is given by
\bea
S_F &=&
\frac{1}{\pi \alpha} \int d\tau {\int_0}^{\pi |\alpha|} d\sigma
    \left[  S^{1a} \partial_+ S^{1a} - \mu S^{1a} \Pi S^{1a} \right.  \nn
    && \hspace{3cm}     \left.  + S^{2a} \partial_- S^{2a} + \mu S^{2a}
    \Pi S^{2a} \right] ,
\eea
where the spinor indices $a$ run from $1$ to $8$. To describe the open string fermions,
we consider the boundary condition which reduces the fermionic degrees of freedom:
\be
   S^{2a} |_{\p \S} = \Om S^{1a} |_{\p \S} ,
\ee
where $\Om$ is a product of the gamma matrices of the world-volume directions of D-brane.
Then, the solutions of equations of motions satisfying the boundary
conditions read
\bea
S^1 (\sigma, \tau) &=& \cos(\mu \tau) S_0
                       -\sin(\mu \tau) \Pi S_0 + \sum_{n \ne 0} C_n
                      \left[ \phi_n ^1 (\sigma, \tau) \Omega S_n
                      +i \rho_n \phi_n ^2 (\sigma, \tau)
                       \Pi S_n \right] , \nn
S^2 (\sigma, \tau) &=& \cos(\mu \tau) \Omega^T S_0 -\sin(\mu \tau)
                       \Pi S_0 + \sum_{n \ne 0} C_n
                      \left[ \phi_n ^1 (\sigma, \tau) \Omega S_n
                      +i \rho_n \phi_n ^2 (\sigma, \tau)
                       \Pi S_n \right]
\eea
where
\bea
\ph_n^1 (\s, \t) &=& e^{-i(\om_n \t -n \s / |\a|)} \quad , \quad
                   \ph_n^2 (\s, \t) = e^{-i(\om_n \t +n \s / |\a|)} , \nn
\rh_n            &=& \frac{\om_n-n/|\a|}{\m}, \nn
c_n              &=&\frac{1}{\sqrt{1+\rh_n^2}} .
\eea

Using the canonical momenta defined by
\be
\l^{Aa} \equiv \frac{\p \c{L}}{\p ( \p_{\t} S^{Aa} )}
      = \frac{i}{\pi \a} S^{Aa} ,
\ee
where $A = 1$ and $2$, the Hamiltonian is given by
\be
\c{H}_F = \frac{i}{\pi \a} \lint{0}{\pi |\a|} d\s \bkb{ S^{1a} \p_\s  S^{1a}
        - S^{2a} \p_\s  S^{2a}
      - \m \bks{ S^{1a} \Pi  S^{1a} - S^{2a} \Pi S^{2a} }}
\ee

At a fixed time $\t=0$, the Hamiltonian is written by
\be
\c{H}_F = - \rm{sign} (\a) 2  \bkb{ i \om_0 S_0 \Om \Pi S_0 -
        \lsum{n \ne 0}{} \om_n S_n S_{-n} }.
\ee
For the non-zero modes ($n>0$), if we define new variables:
\be
S_{-n} \equiv \frac{b_n^+}{2} \quad {\rm and} \quad S_{n} \equiv
               \frac{b_n}{2} ,
\ee
then the Hamiltonian for the non-zero modes is
given by a similar form of the harmonic oscillator
\be
\c{H}_{F,n\ne0} = {\rm sign} (\a) \half \lsum{n>0}{} \om_n
                \bks{b_n^+ b_n -b_n b_n^+} .
\ee
Notice that the Hamiltonian of the fermionic zero modes contains $\Om$, whose form depends
on a configuration of Dp-brane. Hence, in the next section, we will describe, case by case,
spectrum of the fermionic zero modes.

\section{Spectrum of the fermionic zero modes}

In pp-wave background space-time, the symmetry group is given by ${\rm SO(4)} \times {\rm SO(4)}$.
To describe an open string, we have to include Dp-brane. This Dp-brane breaks half supersymmetry
and also break a symmetry group into a appropriate subgroup according to the Dp-brane configuration.
To describe an open string on the Dp-brane world volume, we have to find the representation of an
open string under this subgroup. Moreover, at the low energy scale where the stringy excitation modes
are excluded, the zero mode states of this open string have one-to-one correspondence to
field contents of the SYM theory.

In the bosonic modes of an open string, the existence of a Dp-brane divides coordinates $X^I$ into
two parts: $(p-1)$ Neumann coordinates satisfying $\p_{\s} X^r = 0$, and
$(9-p)$ Dirichlet coordinates satisfying $\p_{\t} X^{r^{\pr}} =0$.
As shown in \eq{bosham}, Dirichlet coordinates have no zero modes, so
the zero point energy is given by \eq{boszero}.

Now let us consider Hamiltonian of the fermionic zero modes
\be
    \c{H}_{F,n=0} = - \rm{sign} (\a) 2 i \om_0 S_0 \Om \Pi S_0 ,
\ee
where $S_0$ are the real eight-component spinors.
From the fact that $\Om \Pi \Om \Pi = -1$,
an eigenvalue of $\Om \Pi$ is given by $+i$ or $-i$.
Suppose that $\psi$ are eigenstates of $\Om \Pi$ with an eigenvalue $+i$ or $-i$,
then the fermionic zero modes $S_0$ can be described by the appropriate linear combinations
of $\psi$ which are complex spinors and depend on the boundary condition.
In the next sections, we will give a full details of these combinations
consistent with the boundary condition.

\begin{table}
\centering  \label{ppwave}
\begin{tabular}  {||c|c|c|c|c||}  \hline
    fermions   & $i \g^{12}$ & $i \g^{34}$   &  $i \g^{56}$  &  $i \g^{78}$      \\ \hline
    $\psi^1$   &  +          &  +            &   +           &   +               \\ \hline
    $\psi^2$   &  +          &  +            &   -           &   -               \\ \hline
    $\psi^3$   &  -          &  -            &   +           &   +               \\ \hline
    $\psi^4$   &  -          &  -            &   -           &   -               \\ \hline
    $\psi^5$   &  +          &  -            &   +           &   -               \\ \hline
    $\psi^6$   &  +          &  -            &   -           &   +               \\ \hline
    $\psi^7$   &  -          &  +            &   +           &   -               \\ \hline
    $\psi^8$   &  -          &  +            &   -           &   +               \\ \hline
\end{tabular}
\caption{The rotation properties of $\psi$ in transverse eight-dimension.}
\end{table}

Without considering the boundary condition, the pp-wave background
has a ${\rm SO(4)} \times {\rm \tilde{SO}(4)}$ Lorentz symmetric
group, which is caused by a constant RR 5-form flux $F_{+1234} =
F_{+5678} = 2\m$. This symmetric group can be described by ${\rm
SU(2)_L} \times {\rm SU(2)_R} \times {\rm \tilde{SU}(2)_L} \times
{\rm \tilde{SU}(2)_R}$, which is a bispinor representation (see
appendix). Now consider ${\rm SO(2)} \times {\rm SO(2)}^{\pr}$,
which are rotations in the $12$ and $34$ and are generated by
$T_{12} \equiv i \g^{12} /2$ and $T_{34} \equiv i \g^{34} /2$
respectively, these rotation generators, $T_{12} $ and $T_{34}$,
are related with the Cartan generators, $J_{3L}$ and $J_{3R}$, of
the first ${\rm SU(2)_L} \times {\rm SU(2)_R}$ symmetric group:
\bea
    J_{3L} &=& \half ( \bks{ T_{12} + T_{34} } )  , \nn
    J_{3R} &=& \half ( \bks{ T_{12} - T_{34} } ) .
\eea
At the same way, other Cartan generators, $\tilde{J}_{3L}$ and $\tilde{J}_{3R}$, of the second
symmetric group ${\rm \tilde{SU}(2)_L} \times {\rm \tilde{SU}(2)_R}$ are given by, in terms of
generators $T_{56}$ and $T_{78}$ of a rotation group ${\rm \tilde{SO}(2)} \times
{\rm \tilde{SO}(2)}^{\pr}$,
\bea
    \tilde{J}_{3L} &=&  \half ( \bks{ T_{56} + T_{78} } )  , \nn
    \tilde{J}_{3R} &=&  \half ( \bks{ T_{56} - T_{78} } ) .
\eea
Now define complex fermions:
\bea
    \psi^{2a-1} &\equiv& \frac{1}{\sqrt{2}} \bks{ S_0^{2a-1} + i S_0^{2a}} , \nn
    \psi^{2a}   &\equiv& \frac{1}{\sqrt{2}} \bks{ S_0^{2a-1} - i S_0^{2a}} ,
\eea
where $a = 1,2,3,4$, and these fermions $\psi$ have a proper eigenvalues of
generators in ${\rm SO(2)} \times {\rm SO(2)}^{\pr} \times {\rm \tilde{SO}(2)} \times
{\rm \tilde{SO}(2)}^{\pr}$, as shown in the table 1.

The Hamiltonian of the fermionic zero modes can be rewritten as
\bea    \label{hamfer}
    \c{H}_{F,n=0} = - {\rm sign} (\a) 2 i \om_0
              &&    \ltb \psi^{2a-1} \bks{\Om \Pi}_{2a-1,2b} \psi^{2b}
                           + \psi^{2a} \bks{\Om \Pi}_{2a,2b-1} \psi^{2b-1}
                    \rte  \nn
              &&    \lte  -i \bkm{
                                   \psi^{2a-1} \bks{\Om \Pi}_{2a-1,2b-1} \psi^{2b-1}
                                 - \psi^{2a} \bks{\Om \Pi}_{2a,2b} \psi^{2b}
                                 }
                    \rtb .
\eea
Notice that if we choose $\psi$ as an eigenstate of $\Om
\Pi$, then $\Om \Pi$ is diagonalized and
the second line in \eq{hamfer} becomes zero due to the Pauli
exclusion principle.

\subsection{For D7-brane}

\begin{table}
\centering  \label{onD7}
\begin{tabular}  {||c|c|c|c||}  \hline
    fermions   &      $(J_{3L}, J_{3R} )$    &  $i \g^{56}$ &  $i \g^{78}$      \\ \hline
    $\psi^1$   &          ( + $\half$ , 0 )          &  +           &   +               \\ \hline
    $\psi^2$   &          ( + $\half$ , 0 )          &  -           &   -               \\ \hline
    $\psi^3$   &          ( - $\half$ , 0 )          &  +           &   +               \\ \hline
    $\psi^4$   &          ( - $\half$ , 0 )          &  -           &   -               \\ \hline
    $\psi^5$   &          ( 0 , +$\half$ )          &  +           &   -               \\ \hline
    $\psi^6$   &          ( 0 , +$\half$ )          &  -           &   +               \\ \hline
    $\psi^7$   &          ( 0 , -$\half$ )          &  +           &   -               \\ \hline
    $\psi^8$   &          ( 0 , -$\half$ )          &  -           &   +               \\ \hline
\end{tabular}
\caption{The eigenvalues of $\psi$ under a symmetric group, ${\rm SU(2)_L} \times
            {\rm SU(2)_R} \times {\rm \tilde{SO}(2)} \times {\rm \tilde{SO}(2)}^{\pr}$.}
\end{table}

In this section, we will consider D7-brane lying in $X^0, X^1, \cdots, X^6,$ and
$X^9$. This D7-brane configuration gives a following symmetric group except light-cone
coordinate directions:
\[
    {\rm SU(2)_L} \times {\rm SU(2)_R} \times \tilde{\rm SO}(2)
                               \times \tilde{\rm SO}(2)^{\pr}   .
\]
Under this symmetric group, fermions $\psi$ transforms, as shown in the table 2.
$\psi^a$ and $\psi^{a+2}$ where $a=1,2$, become a doublet of ${\rm SU(2)_L}$
and $\psi^a$ and $\psi^{a+2}$ where $a=5,6$, appear as a doublet of
${\rm SU(2)_R}$. This implies that a eight-component Majorana-Weyl spinor, $8_s$
in eight-dimensional space is divided by two Weyl spinors with their complex conjugate spinors
in four-dimensional space,
where we use a SO(4) bispinor representation:
\[
    8_s = (2,1) \otimes (\bar{2},1) \otimes (1,2) \otimes (1,\bar{2}) ,
\]
where $(\bar{2},1)$ and $(1,\bar{2})$ are the complex conjugate spinors
of $(2,1)$ and $(1,2)$, respectively. Therefore, if we choose $(2,1)$ and $(1,2)$
as creation operators, $(\bar{2},1)$ and $(1,\bar{2})$ are interpreted as
annihilation operators. These relations can be explicitly shown by writing
a Hamlitonian of fermionic zero modes as a harmonic oscillator form.
To do so, notice that in the case of D7-brane, $\Om \Pi$ is nothing but
$\g^{56}$. Hence, $\psi$ become eigenstates of
$\Om \Pi$ with eigenvalues $\pm i$:
\be
    \Om \Pi \psi^{2a-1} = -i \psi^{2a-1} \quad {\rm and}
        \quad \Om \Pi \psi^{2a} = i \psi^{2a} .
\ee
Therefore, the Hamiltonian of fermionic zero modes on D7-brane is given by
\be
    \c{H}_{F,n=0} = {\rm sign} (\a) 2  \om_0
                  \bkb{ \psi^{2a-1} \psi^{2a} - \psi^{2a} \psi^{2a-1} } .
\ee
Introduce new fermionic variables with the following definitions:
\bea    \label{cafer}
    b_0^{a+} &\equiv& 2 \psi^{2a-1} , \nn
    b_0^a    &\equiv& 2 \psi^{2a} ,
\eea
where $\psi$ in $(2,1)$ and $(1,2)$ are defined as the creation operators
and at the same time, $\psi$ in $(\bar{2},1)$ and $(1,\bar{2})$ are
introduced as the annihilation operators.
Then, the Hamiltonian can be rewritten as a harmonic oscillator form
\be \label{hzfer}
    \c{H}_{F,n=0} = {\rm sign} (\a) \half  \om_0
                  \bkb{ b_0^{a+} b_0^a - b_0^a b_0^{a+} } ,
\ee
where $a$ runs from $1$ to $4$. Therefore, the total Hamiltonian of an open
string ending on Dp-brane is given by
\be     \la{openham}
    \c{H} = {\rm sign} (\a)
            \bkb{
                 \half \sum_{n \ge 0} \omega_n
                       \bks{ {a_n ^I}^+ a_n ^I + a_n ^I {a_n ^I}^+
                             + b_n^{a+} b_n^a -b_n^a b_n^{a+}
                           }
                } .
\ee
In \eq{openham}, notice that for $n=0$ $I$ runs from $1$ to $p-1$, which are
indices of the Dp-brane world-volume coordinates in a light-cone gauge,
and $a$ runs from $1$ to $4$. In the case of non-zero modes ($n \ne 0$),
$I$ and $a$ run from $1$ to $8$.
Therefore, the zero point energy of an open string modes is given by
\be \la{zeroenergy}
    E_0 = {\rm sign} (\alpha) \frac{p-5}{2} \omega_0 ,
\ee
where the zero point energy of non-zero bosonic modes is cancelled by that
of fermionic modes, so only zero modes of an open string give a contribution
to zero point energy. Note that this zero point energy is exactly the same as
that in ref. \ct{sken2}. In the D7-brane case, the zero point energy is given by
${\rm sign} (\alpha) \omega_0$.

At the low energy limit, we expect that the world-volume theory on this D7-brane can be
described by a Super-Yang-Mills theory and all field contents of this theory can be also
obtained from fermionic zero modes of an open string. To obtain field contents of
the D7-brane world-volume theory in the pp-wave background which has a ${\rm SO(4)} \times
\tilde{\rm SO}(2)$ symmetric group, we have to consider the
Fock space consistent with this symmetric group. Here, we pay attention to the SO(4)
representation, so all field contents will be described by fields in four-dimensional
space. In addition, the rest $\tilde{\rm SO}(2)$ symmetric group, which is just
rotation group in $X^5$ and $X^6$ plane, can be considered as a internal symmetric
group. To do so, we divide a spinor indices
$a=1,2,3,4$ into two parts, $\a=1,2$ and $\dot{\a}=1,2$: here, $\a$ and $\dot{\a}$
imply indices of ${\rm SU(2)}_L$ and ${\rm SU(2)}_R$, respectively. Then the Hamiltonian
cab be rewritten as
\be     \la{ohfour}
    \c{H} = {\rm sign} (\a)
            \bkb{
                 \half \sum_{n \ge 0} \omega_n
                       \bks{ {a_n ^I}^+ a_n ^I + a_n ^I {a_n ^I}^+
                             + b_n^{\a+} b_n^{\a} -b_n^{\a} b_n^{\a+}
                             + b_n^{\dot{\a}+} b_n^{\dot{\a}}
                             -b_n^{\dot{\a}} b_n^{\dot{\a}+}
                           }
                } .
\ee
Now, to construct a Fock space we have to choose an appropriate vacuum which is
annihilated by annihilation operators:
\be
    0 = b_0^{\a} |-1,0 \rangle = b_0^{\dot{\a}} |-1,0 \rangle ,
\ee
where $-1$ and $0$ in the state imply the charges of the rotation groups
$\tilde{\rm SO}(2) \times \tilde{\rm SO}(2)^{\pr}$.
Using these fermionic zero modes and the vacuum state,
we can classify the field contents of the world-volume theory,
as shown in the table 3.

\begin{table}
\centering  \label{speconD7}
\begin{tabular}  {c|c|c|c}  \hline
    state               & representation & energy & field      \\ \hline
    $|-1, 0 \rangle$
                        & $(1,1)^{(-1,0)}$          & $\om_0$
                                    &   $A$     \\ \hline
    $b_0^{\a +} |-1, 0 \rangle$
                        & $(2,1)^{(-\half,\half)}$  & $\frac{3}{2} \om_0$
                                    &   $\psi^{\a}$               \\ \hline
    $b_0^{\dot{\a} +} |-1, 0 \rangle$
                        & $(1,2)^{(-\half,-\half)}$ & $\frac{3}{2} \om_0$
                                    &   $\psi^{\dot{\a}}$               \\ \hline
    $b_0^{\a +} b_0^{\b +} |-1, 0 \rangle$
                        & $(1,1)^{(0,1)}$           & $2\om_0$
                                    &   $\phi$               \\ \hline
    $b_0^{\a +} b_0^{\dot{\b} +} |-1, 0 \rangle$
                        & $(2,2)^{(0,0)}$           & $2\om_0$
                                    &   $A^i$               \\ \hline
    $b_0^{\dot{\a} +} b_0^{\dot{\b} +} |-1, 0 \rangle$
                        & $(1,1)^{(0,-1)}$          & $2\om_0$
                                    &   $\bar{\phi}$               \\ \hline
    $b_0^{\a +} b_0^{\dot{\b} +} b_0^{\dot{\g} +} |-1, 0 \rangle$
                        & $(2,1)^{(\half,-\half)}$   & $\frac{5}{2} \om_0$
                                    &   $\bar{\psi}^{\a}$               \\ \hline
    $b_0^{\dot{\a} +} b_0^{\b +} b_0^{\g +} |-1, 0 \rangle$
                        & $(1,2)^{(\half,\half)}$  & $\frac{5}{2} \om_0$
                                    &   $\bar{\psi}^{\dot{\a}}$               \\ \hline
    $b_0^{\a +} b_0^{\b +} b_0^{\dot{\a} +} b_0^{\dot{\b} +} |-1, 0 \rangle$
                        & $(1,1)^{(1,0)}$           & $3\om_0$
                                    &   $\bar{A}$               \\ \hline
\end{tabular}
\caption{The spectrum of an open string ending on D7-brane.}
\end{table}

In table 3, $A$ and $\bar{A}$ are identified with complex scalar fields
having $-1$ and $+1$ charges under the rotation in $X^5$ and $X^6$ plane, which
implies that $A$ and $\bar{A}$ are just the linear combinations of the real
scalar fields, $X^5$ and $X^6$. In six-dimensional space, these two fields together
with $A^i$, which is a gauge field in four-dimensional space, become a six-dimensional
vector field. The other scalar fields, $\ph$ and $\bar{\ph}$ become a singlet under
the ${\rm SO(4)} \times \tilde{\rm SO}(2)$ symmetric group and have $+$ or $-$ charge
of $\tilde{\rm SO}(2)^{\pr}$.

\subsection{For D5-brane}

In this section, we will consider a D5-brane lying in $X^1$,
$X^2$, $X^3$ and $X^8$. The configuration of this D5-brane breaks
a ${\rm SO(4)} \times {\rm \tilde{SO}(4)}$ symmetric group into
${\rm SO(3)} \times {\rm \tilde{SO}(3)} \sim {\rm SU(2)} \times {\rm \tilde{SU}(2)}$
in which ${\rm SO(3)}$ or ${\rm \tilde{SO}(3)}$
is the rotation group in $X^1$, $X^2$, $X^3$ or $X^5$, $X^6$, $X^7$ directions
respectively. Therefore, to find an
effective open string theory on a D5-brane world volume, we have to
investigate the field contents, especially coming from open string zero modes and
transforming under the ${\rm SU(2)} \times {\rm \tilde{SU}(2)}$ symmetric group.
Here, ${\rm SU(2)}$ and ${\rm \tilde{SU}(2)}$ are considered as a diagonal subgroup
of ${\rm SU(2)_L} \times {\rm SU(2)_R}$ and ${\rm \tilde{SU}(2)_L} \times
{\rm \tilde{SU}(2)_R}$ respectively. Here, we choose $X^4$ and $X^8$
as the invariant coordinates under this diagonal subgroup transformation, see Appendix.
Then, the Cartan generators $J_3 = J_{3L} + J_{3R}$ and $\tilde{J}_3
= \tilde{J}_{3L} + \tilde{J}_{3R}$ are proportional to $T_{12}$ and $T_{56}$ respectively.
Since $\g^{34} \g^{78} = - \g^{37} \g^{48}$ and $\Om \Pi = - \g^{48}$, we assume that all
fermions have an eigenvalue of $\g^{48}$. Then all fermions living on a D5-brane have
the eigenvalues of $i \g^{12}$, $i \g^{37}$ and $i \g^{56}$, so an eigenvalue of
$i \g^{48}$ is fixed by the positive chirality of $\g^{1\cdots 8}$ acting on all fermions
living in the transverse eight-dimensional space-time, see table 4.

\begin{table}
\centering  \label{onD5}
\begin{tabular}  {||c|c|c|c|c|c||}  \hline
    fermions   & operators      & $i \g^{12}$ &  $i \g^{37}$  &  $i \g^{56}$  &  $i \g^{48}$      \\ \hline
    $\psi^1$   &$\half b_0^{1+}$&  +          &  +            &   +           &   -               \\ \hline
    $\psi^2$   &$\half b_0^{1 }$&  +          &  +            &   -           &   +               \\ \hline
    $\psi^3$   &$\half b_0^{2+}$&  -          &  -            &   +           &   -               \\ \hline
    $\psi^4$   &$\half b_0^{2 }$&  -          &  -            &   -           &   +               \\ \hline
    $\psi^5$   &$\half b_0^{3 }$&  +          &  -            &   +           &   +               \\ \hline
    $\psi^6$   &$\half b_0^{3+}$&  +          &  -            &   -           &   -               \\ \hline
    $\psi^7$   &$\half b_0^{4 }$&  -          &  +            &   +           &   +               \\ \hline
    $\psi^8$   &$\half b_0^{4+}$&  -          &  +            &   -           &   -               \\ \hline
\end{tabular}
\caption{The rotation properties of $\psi$ ending on a D5-brane.}
\end{table}

As shown in the table 4, note that $\psi^a$ and $\psi^{a+4}$, where $a=1,2,3,4$,
have the same eigenvalues under $i \g^{12}$ and $i \g^{56}$, so they are
distinguished by only the eigenvalue of $i \g^{37}$ or $i \g^{48}$. Here, $\psi^a$ and
$\psi^{a+2}$, where $a=1,2,5,6$, become elements of the doublet of SU(2) and two
doublets, for example $(\psi^1, \psi^3)$ and $(\psi^2, \psi^4)$, are connected by
a ${\rm \tilde{SU}(2)}$ which is an R-symmetry group of N=2 supersymmetry on D5-brane.
To obtain
a Harmonic oscillator form of the hamiltonian, we have to define all fermions
with a positive eigenvalue of $i \g^{48}$ as the creation operators and the rest
are defined as the annihilation operators. Then, the Hamiltonian is written as
\be \   \la{hamonD5}
    \c{H} = {\rm sign} (\a)
            \bkb{
                 \half \sum_{n \ge 0} \omega_n
                       \bks{ {a_n ^I}^+ a_n ^I + a_n ^I {a_n ^I}^+
                             + b_n^{A+} b_n^A -b_n^A b_n^{A+}
                           }
                } ,
\ee
and the zero point energy $E_0$ is given by zero.

Here, since the world-volume theory of D5-brane in the light-cone coordinates has
a SO(3) symmetric group with a $\tilde{\rm SO}(3)$ internal R-symmetric group, we
will find the field contents in terms of three-dimensional space language. In
three-dimensional space, as shown in the table 5., $A^i$ is a vector vector field,
which is composed by three components representing the coordinates, $X^1$, $X^2$
and $X^3$. This vector field, $A^i$ transforms as a vector multiplet under SO(3)
symmetric group. In addition, $A$ and $\bar{A}$ are scalar fields whose linear
combinations imply the scalar fields of $x^5$ and $x^6$ coordinates. $A^0$ implies
$X^7$ field which is a coordinates of target space. This $A^0$ together with $A$
and $\bar{A}$ become a vector multiplet under a $\tilde{\rm SO}(3)$ internal
R-symmetric group. The linear combinations of the other scalar fields, $\ph$
and $\bar{\ph}$, represent $X^4$ and $X^8$. In four-dimensional world-volume
of D5-brane, the vector field $A^i$ in three-dimensional space and a scalar
field $X^8$ become a vector field of D5-brane world-volume theory.

\begin{table}
\centering  \label{speconD5}
\begin{tabular}  {c|c|c|c}  \hline
    state               & representation & energy & field      \\ \hline
    $|0 \rangle$
                        & $1^0$          & $0$
                                    &   $\phi$     \\ \hline
    $b^{\a +} |0 \rangle$
                        & $2^{\half}$  & $\frac{1}{2} \om_0$
                                    &   $\psi^{\a}$               \\ \hline
    $b^{\dot{\a} +} |0 \rangle$
                        & $2^{-\half}$ & $\frac{1}{2} \om_0$
                                    &   $\psi^{\dot{\a}}$               \\ \hline
    $b^{\a +} b^{\b +} |0 \rangle$
                        & $1^1$           & $\om_0$
                                    &   $A$               \\ \hline
    $b^{\a +} b^{\dot{\b} +} |0 \rangle$
                        & $1^0 \bigoplus 3^0$           & $\om_0$
                                    &   $A_0 \bigoplus A^i$               \\ \hline
    $b^{\dot{\a} +} b^{\dot{\b} +} |0 \rangle$
                        & $1^{-1}$          & $\om_0$
                                    &   $\bar{A}$               \\ \hline
    $b^{\a +} b^{\dot{\b} +} b^{\dot{\g} +} |0 \rangle$
                        & $2^{-\half}$   & $\frac{3}{2} \om_0$
                                    &   $\bar{\psi}^{\a}$               \\ \hline
    $b^{\dot{\a} +} b^{\b +} b^{\g +} |0 \rangle$
                        & $2^{\half}$  & $\frac{3}{2} \om_0$
                                    &   $\bar{\psi}^{\dot{\a}}$               \\ \hline
    $b^{\a +} b^{\b +} b^{\dot{\a} +} b^{\dot{\b} +} |0 \rangle$
                        & $1^0$           & $2\om_0$
                                    &   $\bar{\phi}$               \\ \hline
\end{tabular}
\caption{The spectrum of an open string ending on D5-brane.}
\end{table}

\subsection{For D3-brane}

Now, we consider the D3-brane lying in $x^1$ and $x^2$ directions.
This D3-brane configuration breaks the ${\rm SO}(4) \times {\rm SO}(4)$
symmetric group of the pp-wave background into the
${\rm SO}(2) \times {\rm \tilde{SO}}(2) \times {\rm SU}(2) \times
{\rm \tilde{SU}}(2)$ where the first, SO(2) is the Lorentz symmetry group
of the D3-brane world-volume theory in the light-cone gauge. The other group,
${\rm \tilde{SO}}(2) \times {\rm SU}_L (2) \times {\rm SU}_R (2)$ is the internal
symmetric group representing the R-symmetry. To construct the Fock space, we have to
choose the fermion which is an eigenstate of $\Om \Pi$, otherwise there are extra term
in the Hamiltonian, as previously mentioned (see \eq{hamfer}).
Since $\Om \Pi = -\g^{34}$, consider the fermions with some quantum numbers under
the Lorentz group and the internal group, as shown in the table 5.

\begin{table}
\centering  \label{specD7}
\begin{tabular}  {||c|c|c|c|c|c||}  \hline
    fermions   & operators       & $i \g^{12}$ & $i \g^{34}$   &  $i \g^{56}$  &  $i \g^{78}$      \\ \hline
    $\psi^1$   & $ \half b_0^{1+}$ & +          &  -            &   +           &   -               \\ \hline
    $\psi^2$   & $ \half b_0^{1 }$ & -          &  +            &   +           &   -               \\ \hline
    $\psi^3$   & $ \half b_0^{2+}$ & +          &  -            &   -           &   +               \\ \hline
    $\psi^4$   & $ \half b_0^{2 }$ & -          &  +            &   -           &   +               \\ \hline
    $\psi^5$   & $ \half b_0^{3+}$ & -          &  -            &   +           &   +               \\ \hline
    $\psi^6$   & $ \half b_0^{3 }$ & +          &  +            &   +           &   +               \\ \hline
    $\psi^7$   & $ \half b_0^{4+}$ & -          &  -            &   -           &   -               \\ \hline
    $\psi^8$   & $ \half b_0^{4 }$ & +          &  +            &   -           &   -               \\ \hline
\end{tabular}
\caption{The rotation properties of $\psi$ in transverse eight-dimension.}
\end{table}

In the table 6, each $\psi$ is an one-component Weyl fermion which is the fermionic
field in the D3-brane world-volume theory. Now, we define the Cartan generators of
${\rm SU}_L (2)$ and ${\rm SU}_R (2)$ internal symmetric groups as
$J_{3L}$ and $J_{3R}$:
\bea
    J_{3L} &\equiv& \half ( T_{56} + T_{78} ), \nn
    J_{3R} &\equiv& \half ( T_{56} - T_{78} ).
\eea
Then, we find that $(\psi^1,\psi^3)$ and $(\psi^2,\psi^4)$ become doublets under
${\rm SU}_R (2)$ internal group and $(\psi^5,\psi^7)$ and $(\psi^6,\psi^8)$ become
doublets under ${\rm SU}_L (2)$ symmetric group.
If we choose $\psi^a$, where $a=1,3,5,7$,
as creation operators and the rest as annihilation operators, the Hamiltonian
for the open string modes ending on D3-brane can be rewritten by a harmonic oscillator form:
\be \   \la{hamonD3}
    \c{H} = {\rm sign} (\a)
            \bkb{
                 \half \sum_{n \ge 0} \omega_n
                       \bks{ {a_n ^I}^+ a_n ^I + a_n ^I {a_n ^I}^+
                             + b_n^{A+} b_n^A -b_n^A b_n^{A+}
                           }
                } ,
\ee
where in the case of the zero modes ($n=0$), $I$ runs from $1$ to $2$
and $A$ from $1$ to $4$. Therefore, the zero point energy, which is caused by the
different numbers of bosonic and fermionic zero modes, is given by
$E_0 = - {\rm sign} (\alpha) \omega_0$.

Note that the creation operators have $-\half$ quantum number under $T_{34} \equiv
\half i\g^{34}$. Hence, we choose the vacuum which has $+1$ eigenvalue of $T_{34}$ and
represents $|1,0 \rangle$. This vacuum has to be annihilated by the annihilation operators
$b_0^A$ and corresponds to the scalar field like $\frac{1}{\sqrt{2}} (X^3+iX^4)$, which
has $+1$ eigenvalue under ${\rm \tilde{SO}}(2)$. All field contents coming from the
fermionic zero modes are shown in the table 7.

\begin{table}
\centering  \label{speconD7}
\begin{tabular}  {c|c|c|c}  \hline
    state               & (spin) representation & energy & field      \\ \hline
    $|1, 0 \rangle$
                        & $(0)^{(1,0,0)}$ & $- \om_0$ &   $A$     \\ \hline
    $b_0^{A +} |1, 0 \rangle$
                        & $(\half)^{(\half,\half,-\half)} \bigoplus
                        (\half)^{(\half,-\half,\half)} $
                         & $- \half \om_0$
                                    &   $\psi^{A}$               \\
                        & $\bigoplus (-\half)^{(\half,\half,\half)}
                        \bigoplus (-\half)^{(\half,-\half,-\half)}$
                         &          &  \\ \hline
                        & $
                        (1)^{(0,0,0)} \bigoplus (-1)^{(0,0,0)}$
                         &
                                    &   $A^i$               \\
  $b_0^{A +} b_0^{B +} |1, 0 \rangle$ & $\bigoplus (0)^{(0,1,0)} \bigoplus (0)^{(0,-1,0)}$
                         & $0$
                        & $\bigoplus \ph^1 \bigoplus \bar{\ph}^1$
                            \\
                        & $\bigoplus (0)^{(0,0,1)} \bigoplus (0)^{(0,0,-1)} $ &
                        & $\bigoplus \ph^2 \bigoplus \bar{\ph}^2$ \\ \hline
    $b_0^{A +} b_0^{B +} b_0^{C +} |1, 0 \rangle$
                        & $(-\half)^{(-\half,-\half,\half)} \bigoplus
                        (-\half)^{(-\half,\half,-\half)} $
                        & $\half \om_0$ &   $\bar{\psi}^{A}$  \\
                        & $\bigoplus (\half)^{(-\half,-\half,-\half)}
                        \bigoplus (\half)^{(-\half,\half,\half)}$ & &  \\ \hline
    $b_0^{A +} b_0^{B +} b_0^{C +} b_0^{D +} |1, 0 \rangle$
                        & $(0)^{(-1,0,0)}$ & $\om_0$
                                    &   $\bar{A}$               \\ \hline
\end{tabular}
\caption{The spectrum of an open string ending on D3-brane.}
\end{table}

In the table 7, $A$ and $\bar{A}$ are the linear combinations of the $X^3$ and $X^4$
scalar fields. Moreover, $(\ph^1, \bar{\ph}^1)$ and $(\ph^2, \bar{\ph}^2)$ are
also the linear combinations of $(X^5,X^6)$ and $(X^7,X^8)$, respectively.

\section{Discussion}

By many authors, the open string theory and their superalgebra on the D-brane
in the pp-wave background are studied.
In this paper, we have exactly shown that the field contents of the SYM theory with
the non-trivial background are obtained from the open string theory in the
pp-wave background. Note that the world-volume theory is described by a
low dimensional representation, which arises due to the non-trivial symmetric group.
Like the Ref. \ct{met} where SYM theory as a world-volume theory of D3-brane in the pp-wave
background is studied, it has been believed that the world-volume theory on the D-brane in
the pp-wave background is SYM theory. However, although it is not clear how
in the GS formalism
we can obtain the world-volume theory such as the Dirac-Born-Infeld action in the
flat space background, the result of this paper gives some evidences that the world-volume
theory of D-brane in the pp-wave can be described by a SYM theory.

In the case of the closed string field theory, in the Ref. \ct{spradlin2} the interaction
Hamiltonian of the three closed strings was obtained and in the Ref. \ct{kiem1} it was
shown that this interaction Hamiltonian is a correct one compared with the supergravity
calculation. In the open string field theory case, the interaction Hamiltonian was obtained
using the similar method of the closed string field theory. To believe that this
interaction Hamiltonian is correct, we have to compare with the SYM theory result.
So it is very interesting to compare the open string field theory result with that
of SYM theory, we expect that the result of this paper has a crucial role in studying
that topic.

\vspace{2cm}
\noindent {\bf Acknowledgement:} We thank H. S. Yang and B. Lee for helpful discussion.
This work was supported by Korea Research Foundation
Grant. (KRF-2002-037-C00007).

\appendix

\section*{Appendix}

\section{Spinor representation}

In this section, we will give a summary for a symmetry group, especially SO(4)
in four-dimensional Euclidean space which was used in constructing
spinors on D5-brane world-volume from type IIB string theory, see also a ref \ct{oz}
for more details. In four-dimensional
Euclidean space, the symmetry group leaving the metric invariant is O(4).
Let ${\bf V}$ denote a four component real vector with entries $V^I$. Then,
this O(4) symmetry group is composed of SO(4) symmetry group, which is a
{\it proper} subgroup of O(4), and a parity transformation:
reflection of an odd number of directions, e.g. $V^I \to -V^I$.

Here, we concentrate on a {\it proper} subgroup, SO(4). A vector representation of
SO(4) is characterized by invariance of a quadratic form under a group of linear
transformation:
\be     \la{vecrep}
    \bk{{\bf V}, {\bf V}} \equiv G_{IJ} V^I V^J ,
\ee
where $I$ runs from $0$ to $3$.
Since in a Euclidean space, without loss of generality, we can set the metric, $G_{IJ}$
to be equal to an identity matrix $\d_{IJ}$, so $\bk{{\bf V}, {\bf V}}$ is just the sum
of the squares of $V^I$. Now define
\bea    \la{bi-spinrep}
    {\c{V}^{\a}}_{\dot{\a}} &\equiv& {(V^0 \id  + i V^i \s^I )^{\a}}_{\dot{\a}} \nn
                            &=& \left( \ba {rr}
                                               V^0 + V^3     &    V^2 + V^1     \\
                                              -V^2 + V^1     &    V^0 - V^3     \\
                                       \ea
                                 \right)     ,
\eea
where $i$ runs from $1$ to $3$.
Then, \eq{vecrep} can be rewritten as
\be
    \det \c{V} = \bk{{\bf V}, {\bf V}} ,
\ee
so a component of a four vector, $V^I$, is given by
\be     \la{relvecspin}
    V^0 = \half \Tr{\bks{\c{V}}} , \quad V^i = - \frac{i}{2} \Tr{\bks{\c{V} \s^i } } .
\ee
Due to a reality condition of $V^I$, a complex conjugate of $\c{V}$ is given by
\be     \la{realcond}
    \bks{{\c{V}^{\a}}_{\dot{\a}}}^* = {(\s^2)^{\a}}_{\b} {\c{V}^{\b}}_{\dot{\b}}
                                        {(\s^2)^{\dot{\b}}}_{\dot{\a}} .
\ee
Now let us consider a linear transformation:
\be     \la{lintransf}
    \c{V} \goto U \c{V} U^{\pr} .
\ee
Imposing a reality condition, \eq{realcond}, and an invariance of the above quadratic form, \eq{vecrep},
gives two constraints for a linear transformation:
\bea
    \det U      &=& 1 , \nn
    \s^2 U \s^2 &=& U^* ,
\eea
and there are also the same constraints for $U^{\pr}$. These constraints imply that $U$ and $U^{\pr}$
are just elements of SU(2) group. Note that there is no condition to relate $U$ and $U^{\pr}$.
To distinguish these two SU(2) groups, we denote $U$ and $U^{\pr}$ as elements of ${\rm SU(2)}_L$
and ${\rm SU(2)}_R$, respectively. As a result, a vector of SO(4) can be described by
a bi-spinor form as \eq{bi-spinrep}, which realizes the (2,2) representation of
${\rm SU(2)}_L \times {\rm SU(2)}_R$. Using \eq{vecrep} and \eq{bi-spinrep},
we can easily convert a vector representation of SO(4) to a bi-spinor representation, or vice versa.

This SO(4) symmetry group has several subgroups. Clearly, SO(3) group is an subgroup of SO(4) symmetry
group, which leaves $V^0$ invariant. Equivalently, this subgroup can be realized as a diagonal
subgroup of ${\rm SU(2)}_L \times {\rm SU(2)}_R$. The invariance of $V^0$ under a linear transformation
\eq{lintransf}, gives the following relation
\be     \la{diasubgr}
    U U^{\pr} = \id ,
\ee
which characterizes a diagonal subgroup of ${\rm SU(2)}_L \times {\rm SU(2)}_R$. Usually,
since an element of a Lie group can be written as the following form
\be
    U = e^{i \L_i \s^i} ,
\ee
the above relation \eq{diasubgr}, means that $U^{\pr}$ must be given by
\be
    U^{\pr} = e^{-i \L_i \s^i} .
\ee
Now, define the Cartan generators, $J_{3L} \equiv \s^3 /2$ in ${\rm SU(2)}_L$
and $J_{3R} \equiv \s^3 /2$ in ${\rm SU(2)}_R$.
Due to the equivalence of $J_{3L}$ and $J_{3R}$ in a diagonal subgroup,
an eigenstate of $J_{3L}$ is also that of $J_{3R}$.
Suppose that there is a fermion which is an eigenstate of
$J_{3L}$, then this fermion can be characterized by a quantum number of
$J_{3L}$ and $J_{3R}$. However, since $J_{3L} - J_{3R} = 0$ a fermion is charcterized
by a quantum number of $J_3 \equiv J_{3L} + J_{3R}$ only.
Finally, to find a appropriate fermion described by spinor representation of a SO(3) subgroup,
we have to choose a spinor which is an eigenstate of $J_3$.

\newpage


\nc{\np}[3]{Nucl. Phys. {\bf B#1}, #2 (#3)}

\nc{\pl}[3]{Phys. Lett. {\bf B#1}, #2 (#3)}

\nc{\prl}[3]{Phys. Rev. Lett. {\bf #1}, #2 (#3)}

\nc{\prd}[3]{Phys. Rev. {\bf D#1}, #2 (#3)}

\nc{\ap}[3]{Ann. Phys. {\bf #1}, #2 (#3)}

\nc{\prep}[3]{Phys. Rep. {\bf #1}, #2 (#3)}

\nc{\ptp}[3]{Prog. Theor. Phys. {\bf #1}, #2 (#3)}

\nc{\rmp}[3]{Rev. Mod. Phys. {\bf #1}, #2 (#3)}

\nc{\cmp}[3]{Comm. Math. Phys. {\bf #1}, #2 (#3)}

\nc{\mpl}[3]{Mod. Phys. Lett. {\bf #1}, #2 (#3)}

\nc{\cqg}[3]{Class. Quant. Grav. {\bf #1}, #2 (#3)}

\nc{\jhep}[3]{J. High Energy Phys. {\bf #1}, #2 (#3)}

\nc{\hep}[1]{{\tt hep-th/{#1}}}



\begin{thebibliography}{99}



\bibitem{blau} M. Blau, J. Figueroa-O'Farrill, C. Hull, and
G. Papadopoulos, \jhep {01}{047}{2002}, \hep{0110242}; \cqg
{19}{L87}{2002}, \hep{0201081}.

\bibitem{met1} R. R. Metsaev, \np {625}{70}{2002}, \hep{0112044}.

\bibitem{met2} R. R. Metsaev and A. A. Tseytlin, \prd {65}{126004}{2002},
\hep{0202109}.

\bibitem{bmn} D. Berenstein, J. Maldacena, H. Nastase,
\jhep {0204} {013} {2002}.

\bibitem{kris} C. Kristjansen, J. Plefka, G. W. Semenoff, and M.
Staudacher, \np {643}{3}{2002}, \hep{0205033}.

\bibitem{gross} D. J. Gross, A. Mikhailov, and R. Roiban,
\ap {30}{31}{2002}, \hep{0205066}.

\bibitem{constanble} N. R. Constable, D. Z. Freedman, M. Headrick,
S. Minwalla, L. Motl, A. Postnikov, and W. Skiba, \jhep
{07}{017}{2002}, \hep{0205089}.

\bibitem{chu} C.-S. Chu, V. V. Khoze, and G. Travaglini, \jhep
{06}{011}{2002}, \hep{0206005}.

\bibitem{santa} A. Santambrogio and D. Zanon, \pl {545}{425}{2002}, \hep{0206079}.

\bibitem{beisert} N. Beisert, C. Kristjansen, J. Plefka, G. W. Semenoff,
and M. Staudacher, \np {650}{125}{2003}, \hep{0208178}.

\bibitem{const} N. R. Constable, D. Z. Freedman, M. Headrick, and
S. Minwalla, \jhep {10}{068}{2002}, \hep{0209002}.

\bibitem{gursoy} U. G\"ursoy, \hep{0208041}; \hep{0212118}.

\bibitem{kiem1} Y. Kiem, Y. Kim, S. Lee, and J. Park, \np
{642}{389}{2002}, \hep{0205279}.

\bibitem{huang} M.-x. Huang, \pl {542}{255}{2002}, \hep{0205311}.

\bibitem{lee1} P. Lee, S. Moriyama, and J. Park, \prd
{66}{085021}{2002}, \hep{0206065}.

\bibitem{spradlin2} M. Spradlin and A. Volovich,
\jhep {01}{036}{2003}, \hep{0206073}

\bibitem{chu1} C.-S. Chu, V. V. Khoze, M. Petrini, R. Russo, and A.
Tanzini, \hep{0208148}.

\bibitem{pankiewicz1} A. Pankiewicz, \jhep {09}{056}{2002},
\hep{0208209}.

\bibitem{pankiewicz2} A. Pankiewicz and B. Stefa\'nski, Jr,
\np {657}{79}{2003}, \hep{0210246}.

\bibitem{kiem2} Y. Kiem, Y. Kim, J. Park, and C. Ryou, \jhep
{01}{026}{2003}, \hep{0211217}.

\bibitem{spradlin} M. Spradlin and A. Volovich, \prd {66} {086004} {2002}.

\bibitem{gs1} M. B. Green and J. H. Schwarz, L. Brink, \np {219} {437} {1983}.

\bibitem{gs2} M. B. Green and J. H. Schwarz  \np {243} {285} {1984}.

\bibitem{gs3} M. B. Green and J. H. Schwarz, \np {243} {475} {1984}.

\bibitem{ck} B. Chandrasekhar and A. Kumar,
\jhep{0306}{001}{2003};  B. Stefanski Jr, hep-th/0304114.

\bibitem{lam} N.D. Lambert and P.C. West, \pl {459} {515} {1999}.


\bibitem{dab} A. Dabholkar and S. Parvizi, \np {641} {223} {2002}.


\bibitem{billo} M. Billo' and I. Pesando, \pl {536}{121}{2002},
\hep{0203028}.


\bibitem{dabholkar} A. Dabholkar and S. Parvizi, \np
{641}{223}{2002}, \hep{0203231}.


\bibitem{sken-tayl} K. Skenderis and M. Taylor, \jhep
{06}{025}{2002}, \hep{0205054}.


\bibitem{bergman} O. Bergman, M. R. Gaberdiel, and M. B. Green, \hep{0205183}.


\bibitem{bain} P. Bain, K. Peeters, and M. Zamaklar,
\prd {67}{066001}{2003}, \hep{0208038}; P.Bain, P.Meessen, M.Zamaklar,
\cqg {20}{913}(2003).

\bibitem{hikida} Y. Hikida and S. Yamaguchi,
\jhep {01}{072}{2003}, \hep{0210262}.


\bibitem{sken1} K. Skenderis and M. Taylor, \hep{0211011}.


\bibitem{sken2} K. Skenderis and M. Taylor, \hep{0212184}.


\bibitem{gaberdiel} M. R. Gaberdiel and M. B. Green,
\hep{0211122}; M. R. Gaberdiel, M. B. Green, S. Schafer-Nameki and
A. Sinha, hep-th/0306056.


\bibitem{hsy1} J. Kim, B. Lee and H. S. Yang, \prd {68} {026004} {2003}.

\bibitem{hsy2} K. Cha, B. Lee and H. Yang, \hep{0307146}.

\bibitem{ohta} N. Ohta, K. L. Panigrahi and S. Jhingan, \hep{0306186}.

\bibitem{park1} S. Hyun, J. Park and S. Yi, \jhep{0303}{004}{2003}, \jhep{0211}{001}{2002}.

\bibitem{met} R.R. Metsaev, \np{655}{3}{2003}.

\bibitem{oz} Y. E. Cheung, Yaron Oz and Zheng Yin, \hep{0211147}.

\end{thebibliography}
\end{document}